\documentclass[useAMS,usenatbib, times]{mn2e}

\long\def\symbolfootnote[#1]#2{\begingroup%
\def\thefootnote{\fnsymbol{footnote}}\footnote[#1]{#2}\endgroup} 
\def\aj{AJ}%          % Astronomical Journal
\def\apj{ApJ}%          % Astrophysical Journal
\def\mnras{MNRAS}%          % Monthly Notices of the RAS
\def\pasp{PASP}%          % Publications of the ASP
\def\pasj{PASJ}%          % Publications of the ASJ
\def\kms{\text{km} \ \text{s}^{-1}}
 
\def\nbody{$N$-body\ }

\usepackage{amssymb,latexsym,graphicx,natbib,eufrak,times,amsmath,verbatim}
\defcitealias{2003MNRAS.340..227B}{BM03}

\usepackage{lscape}

\usepackage{epstopdf}
\usepackage{color,subfigure,rotating}

\usepackage[english]{babel}
\usepackage{nicefrac,soul}

\usepackage{makecell}

\title[Neutron star kicks and star cluster survival]
  {Neutron star natal kicks and the long-term survival of star clusters}
\author[F. Contenta et al.]
  {Filippo Contenta$^1$,
  Anna Lisa Varri$^2$,
  Douglas C. Heggie$^2$
\\
$^1$  Department of Physics, University of Surrey, Guildford GU2 7XH, UK\\
$^2$  School of Mathematics and Maxwell Institute for Mathematical Sciences, University of Edinburgh, Kings Buildings, Edinburgh EH9 3FD, UK\\
}
\def\LaTeX{L\kern-.36em\raise.3ex\hbox{a}\kern-.15em
    T\kern-.1667em\lower.7ex\hbox{E}\kern-.125emX}

\begin{document}      

\date{Accepted 2015 February 08. Received 2015 February 03; in original form 2014 December 22}

\pagerange{\pageref{firstpage}--\pageref{lastpage}} \pubyear{2015}

\maketitle

\label{firstpage}

\begin{abstract}
We investigate the dynamical evolution of a star cluster in an external tidal field by using \nbody simulations, with focus on the effects of the presence or absence of neutron star natal velocity kicks. We show that, even if neutron stars typically represent less than 2\% of the total bound mass of a star cluster, their primordial kinematic properties may affect the lifetime of the system by up to almost a factor of four. We interpret this result in the light of two known modes of star cluster dissolution, dominated by either early stellar evolution mass loss or two-body relaxation. The competition between these effects shapes the mass loss profile of star clusters, which may either dissolve abruptly (``jumping''), in the pre-core-collapse phase, or gradually (``skiing''), after having reached core collapse.  
\end{abstract}
\begin{keywords}
methods: numerical, stellar dynamics --
globular clusters: general 
\end{keywords}

\section{Introduction}

The rate at which star clusters lose mass has been one of the enduring problems of stellar dynamics.  In one of the earliest results, \citet{amb38} already highlighted the role of relaxation. The landmark survey of \citet{1990ApJ...351..121C} added  a mass spectrum, stellar evolution and a tidal boundary, and also revealed the importance of the initial structure of the star cluster. 
But several other factors also influence the lifetime of star clusters, including the binary population (e.g. \citealt{2009PASJ...61..721T}), the form of the Galactic orbit (e.g. \citealt{2003MNRAS.340..227B}), the form of the Galactic potential and tidal shocking (e.g. \citealt{1997ApJ...474..223G}), and the crossing time scale \citep{2013ApJ...778..118W}.

In this Letter we add one more influence: natal kicks of neutron stars (NS). Though neutron stars may account for less than 2\% of the cluster by mass, we find, astonishingly, that the presence or absence of kicks may change the lifetime of a star cluster by almost a factor of four.  Though the existence of natal kicks of neutron stars is not in doubt, their distribution and dispersion are difficult to establish (see, for example, \citealt{2005ASPC..328..327P}).  

In order to isolate the effect of this one factor we consider models from another landmark survey of the evolution of star clusters: that
by \citet[][hereafter BM03]{2003MNRAS.340..227B}.  As it happens, they imposed no natal kicks on neutron stars, and it was the attempt to reproduce some of their results that led to our discovery. Indeed their principal models, which begin with a King profile with $W_0=5$, evolve very differently, both qualitatively and quantitatively, if natal kicks are applied. 

The particular models we considered are described in the following
section, while Sect.~\ref{Results} presents our results in some
detail, including some information on 
core collapse and  
mass segregation.  The final section summarises our conclusions, and attempts to interpret them in the context of other recent work.

\section{Description of the Runs}\label{Simulation}

We simulate the evolution of a globular cluster as in
\citetalias{2003MNRAS.340..227B}, but using \textsc{NBODY6}
\citep{2012MNRAS.424..545N}. We have performed a survey of simulations
in an accelerating, non-rotating frame, using a number of particles
between $N=8192$ and $N=131072$, a Kroupa IMF
\citep{2001MNRAS.322..231K}, with the mass of the stars between $0.1
\ M_\odot$ and $15 \ M_\odot$ (resulting in a theoretical mean mass
$\left \langle m \right \rangle=0.547 \ M_\odot$), and metallicity
$Z=0.001$. Natal kicks, when they were applied, had a Maxwellian
distribution with $\sigma = 190 \ \kms$ \citep[see eq.~3 in][]{1997MNRAS.291..569H}.

\begin{table*}
\begin{minipage}{126mm}
\caption{\nbody simulation properties}
\label{parameter}
\resizebox{\linewidth}{!}{

\begin{tabular}{@{}lccccccccccc}
 
%\hline
\Xhline{2\arrayrulewidth}
\\[-2ex]  
Model & $N$ & $W_0$ & $e$ & $M_0$ & $r_h$ & $r_J$ & $T_{diss}$ & ${T_{diss}^{BM}}$ & $T_{cc}$&$T_{cc}^{BM}$\\
 \ & \ & \ & \ & $\left[M_\odot\right]$ & $\left[pc\right]$ &
 $\left[pc\right]$ & [Myr] & [Myr] & [Myr]&[Myr]\\
%\hline
\\[-2ex]
\Xhline{2\arrayrulewidth}
\\[-2ex]
8kK & 8192 & 5.0 & 0.0 & 4497.3 & 4.53 & 24.35 & 2426  & - & 2666 & -\\
16kK & 16384 & 5.0 & 0.0 & 8990.7 & 5.73 & 30.67 & 2816  & - & - & -\\
32kK & 32768 & 5.0 & 0.0 & 18419.2 & 7.23 & 38.96 & 3669  & - & - & -\\
64kK & 65536 & 5.0 & 0.0 & 36183.1 & 9.10 & 48.79 & 4516  & - & - & -\\
128kK* & 131072 & 5.0 & 0.0 & 71422.0 & 11.46 & 61.21 & 5927  & - & - & -\\
%\hline
\\[-2ex]
\Xhline{1.5\arrayrulewidth}
\\[-2ex]
8kN & 8192 & 5.0 & 0.0 & 4497.3 & 4.53 & 24.35 & 4137 & 4149 & 3142 & 3329\\
16kN & 16384 & 5.0 & 0.0 & 8990.7 & 5.73 & 30.67 & 5932 & 6348 & 4810 & 5062\\
32kN & 32768 & 5.0 & 0.0 & 18419.2 & 7.23 & 38.96 & 9384 & 9696 & 7788 & 8412\\
64kN & 65536 & 5.0 & 0.0 & 36183.1 & 9.10 & 48.79 & 14414 & 15197 & 12375 & 13193\\
128kN & 131072 & 5.0 & 0.0 & 71659.0 & 11.46 & 61.27 & 22707 & 23769 & 20307 & 21339\\
%\hline
\\[-2ex]
\Xhline{2\arrayrulewidth}
\\[-2ex]
128kKe & 131072 & 5.0 & 0.5 & 71453.0 & 5.50 & 29.43 & 5479 & - & 6859 & -\\
128kNe & 131072 & 5.0 & 0.5 & 71453.0 & 5.50 & 29.43 & 11254 & 11675 & 8952 & 9332\\
128kK7 & 131072 & 7.0 & 0.0 & 71780.9 & 7.14 & 61.31 & 18369 & - & 18267 & - \\
128kN7 & 131072 & 7.0 & 0.0 & 71780.9 & 7.14 & 61.31 & 24494 & 25506 & 11886 & 12620\\
%hline
\\[-2ex]
\Xhline{2\arrayrulewidth}  
\\[-2ex]
\end{tabular}}

\medskip
%% mimicking deluxetable style
Note. --- The capital letter in the model label indicates if the model
is characterized by the presence (\#K, e.g. 128kK) or the absence
(\#N, e.g. 128kN) of NS initial kicks. The star (*) denotes a model
for which two different numerical realizations have been evolved; the
values
are the average of those
for the two 
simulations.
\end{minipage}
\end{table*}

In our simulations, the cluster is in a circular orbit, or in an elliptical orbit with eccentricity $e=0.5$, in a logarithmic Galactic potential $\phi={V_G}^2\ln(R_G)$, where ${V_G}$ is the circular velocity and $R_G$ is the Galactocentric distance. For the majority of our runs we have used a Roche-lobe filling \citet{1966AJ.....71...64K} model with $W_0=5$ as initial condition.  Additional simulations have been
performed by increasing the initial concentration of the King profile
($W_0=7$). The clusters start at a Galactic radius of $8.5$ kpc, with
an initial velocity of $220 \ \text{km} \ \text{s}^{-1}$ (in the
circular case); in the elliptical case the apogalacticon is at $8.5$
kpc and the initial speed there is reduced appropriately, while the
size of the cluster is determined by assuming a Roche-lobe filling
condition 
at perigalacticon. The initial conditions for all the simulations have been generated using M{\sevensize C}L{\sevensize USTER} \citep{2011MNRAS.417.2300K}. 

The tidal radius of the cluster was defined as the Jacobi radius
\begin{equation}
r_J=\left(\frac{G M}{2{V_G}^2}\right)^{\nicefrac{1}{3}}{R_G^{\nicefrac{2}{3}}}, \qquad  
\end{equation}
where $M$ is the ``bound'' cluster mass.  The quantities $M$ and $r_J$ were determined self-consistently and iteratively by first assuming that all stars are still bound and calculating the tidal radius with this formula. In a second step, we calculated the mass of all stars inside
$r_J$ relative to the density centre of all stars, and used it to obtain a new estimate for $r_J$. This method was repeated until convergence. 
Escapers were not removed from the simulations.

The properties of the simulations are presented in
Table~\ref{parameter}.  The significance of the model label is stated
in the note to the Table.  Column 4 gives the orbital eccentricity;
columns 5, 6 and 7 are the initial values of the total bound mass, the
half-mass radius and the Jacobi radius, respectively. Column 8 gives
the dissolution time, which, following
\citetalias{2003MNRAS.340..227B}, is defined as the time when $95\%$
of the mass was lost from the cluster, while column 10 gives the
core-collapse time.  The corresponding quantities
from \citetalias{2003MNRAS.340..227B} are reported in columns 9 and
11, respectively.  In our analysis, the moment of core collapse $T_{cc}$ has been determined by inspecting the time evolution of the core radius and of the innermost lagrangian radius enclosing $1\%$ of the total mass.    

\section{Results}\label{Results}

\subsection{Lifetime and main properties of the models}

\begin{figure}
\includegraphics[trim=0 20 0 50, clip, width=0.48\textwidth]{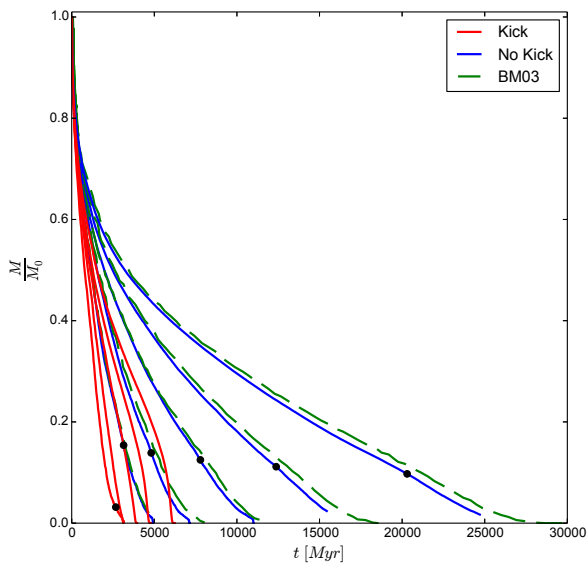}
\caption{Time evolution of the fraction of bound mass (normalized to
  the initial value) of models with initial concentration $W_0=5$, on
  circular orbits. The models are characterized by different number of
  particles and by the presence (red lines; right to left: $128$kK,
  $64$kK, $32$kK, $16$kK and $8$kK) or the absence (blue lines; right
  to left: $128$kN, $64$kN, $32$kN, $16$kN and $8$kN) of NS kicks.  The black dots mark when core collapse occurs. The
  corresponding models studied by \citetalias{2003MNRAS.340..227B},
  i.e. without NS kicks, are also shown (green dashed lines; right to
  left: $128$k, $64$k, $32$k, $16$k and $8$k); this data was retrieved
  by means of the data extraction tool Dexter. 
  }\label{fig:MboverM0_128k}
\end{figure}

The main result of our investigation is that the presence or the absence of NS natal velocity kicks can affect significantly the lifetime of star clusters, up to almost a factor of four. This striking result is illustrated by a series of ``reference models'' (with $W_0=5$, $e=0$, and N = $128$k, $64$k, $32$k, $16$k, and $8$k), with or without NS velocity kicks; the time evolution of the bound mass of these models is presented in Fig.~\ref{fig:MboverM0_128k}. The difference in the behaviour of the models with or without NS kicks starts early in their evolution ($M/M_0 \approx 0.8$) and leads to a dramatic contrast in the slope of the graph at the final stages of evolution ($M/M_0<0.2$).

An important aspect of the very rapid dissolution of the models with NS kicks is that, in almost all cases, they fail to reach core collapse during their evolution, as opposed to the models without NS kicks, which show signatures of core collapse at a time corresponding to $0.1 < M/M_0 < 0.2$. The only exception is given by model $8$kK, which reaches core collapse at the very late stages of evolution, $240$ Myr after the formal dissolution time (see Tab.~\ref{parameter}, row~1, and the corresponding black dot in Fig.~\ref{fig:MboverM0_128k}). The fact that it reaches core collapse at all, while larger models do not, is attributable to its short initial relaxation time.

We have also considered two additional pairs of models, as
representative cases of the regime of high initial concentration
($W_0=7$; models $128$kK7 and $128$kN7) and of the evolution of a star
cluster on an elliptic orbit ($e=0.5$; models $128$kKe and
$128$kNe). Here the effects of the presence of NS kicks on the star
cluster lifetime are less severe compared to those on the ``reference
models'', but they are still significant (see Fig.~\ref{fig:MboverM0_ECC_W7}). Both models
$128$kK7 and $128$kKe reach core collapse, although at a very late
stage of evolution.  Of the systems without NS kicks, model $128$kNe reaches core collapse at a mass comparable to that of ``reference models'' without NS kicks, while model $128$kN7 has the largest mass at $T_{cc}$ of all the models in our survey which reach core collapse; such a result is not surprising, given its initial concentration.

\begin{figure}
\includegraphics[trim=0 20 0 50, clip, width=0.48\textwidth]{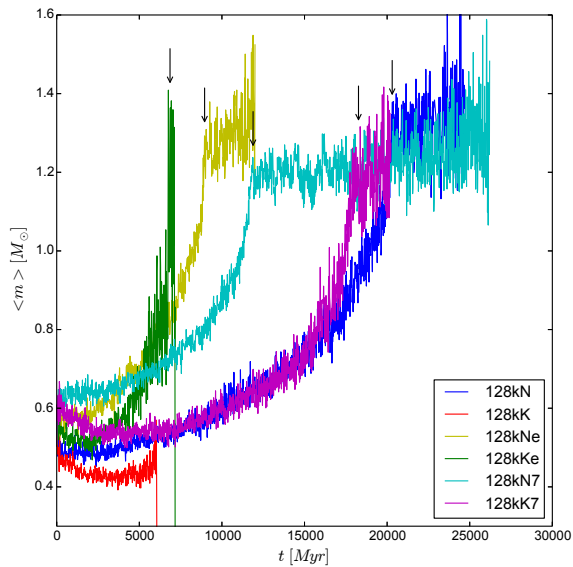}
\caption{Evolution of the mean mass of the stars in the innermost 
  langrangian shell, containing 1\% of the bound mass, evaluated 
for
  all models with N=$128$k. The vertical arrows mark the moment of
  core collapse (in the five models which exhibit core collapse).}\label{fig:mmean}
\end{figure}

Another useful diagnostic of the differences between models with and without NS kicks is provided by the mean mass of stars in the
 innermost lagrangian shell, enclosing $1\%$ of the total bound mass
 of a system. Its time evolution is illustrated in
 Fig.~\ref{fig:mmean}, for all models in our survey with $N=128$k
 particles. In almost all cases, the mean mass in the innermost shell
 initially shows a decrease, which is due to the early
   evolution and escape
of massive stars;  
as expected, this effect is more pronounced for systems with NS kicks. Nonetheless, after only a few Gyr, the value of the central mean mass starts to increase, reflecting the process of mass segregation. For models that reach core collapse, the mean mass in the final stages of evolution falls within the range $1.2 \le \left<m\right>
 \le 1.4 M_{\odot}$, which indicates the dominance of neutron stars in the central regions of the system. Not surprisingly, the rapidly dissolving model $128$kK (Fig.~\ref{fig:mmean}, red line) shows a final mean mass which is comparable to the initial value. 
 
\subsection{Detailed comparison with Baumgardt \& Makino (2003)}

\begin{figure}
\includegraphics[trim=0 20 0 50, clip, width=0.48\textwidth]{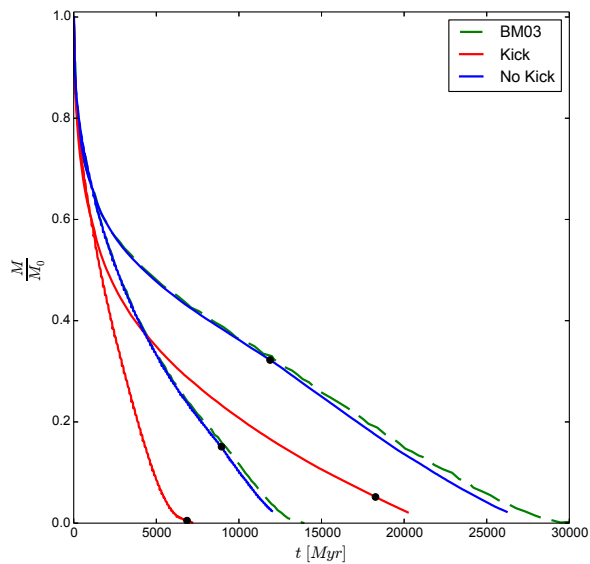}
\caption{Time evolution of the fraction of bound mass of models with
  (i) $W_0=5$, on elliptic orbits; (ii) $W_0=7$, on circular
  orbits. As in Fig.~\ref{fig:MboverM0_128k}, models with NS kicks are
  denoted by red lines (right to left: $128$kK7 and $128$kKe), and
  without NS kicks by blue lines (right to left: $128$kN7 and
  $128$kNe). Dashed green lines show the corresponding
    models (without kicks) from BM03.
}\label{fig:MboverM0_ECC_W7}
\end{figure}

Despite our best efforts in reproducing the initial conditions and the
numerical set-up described by \citetalias{2003MNRAS.340..227B}, we
note that there are still  non-negligible discrepancies between our
models without NS natal kicks and the corresponding ones in their
original investigation (see Table~\ref{parameter} and
Figs.~\ref{fig:MboverM0_128k} and \ref{fig:MboverM0_ECC_W7}). We have attempted to identify the main reasons for these discrepancies in the intrinsic differences
between the \nbody codes used to perform the simulations, and in particular slightly different stellar evolution prescriptions.

We performed our simulations by using the GPU version of
\textsc{NBODY6} \citep{2012MNRAS.424..545N}, 
while
\citetalias{2003MNRAS.340..227B} used the public \textsc{GRAPE-6}
version of \textsc{NBODY4} \citep{1999PASP..111.1333A}. The
  latter treats components of binaries as single stars, without
  collisions or exchange of mass, and the resulting differences  might partially explain the increasing discrepancy after core collapse for the models depicted in Fig.~\ref{fig:MboverM0_ECC_W7}, because of the increase in the number of binaries at this time. 
Moreover, \citetalias{2003MNRAS.340..227B} used a prescription for
the properties of stellar remnants by \citet{2000MNRAS.315..543H}, 
while in \textsc{NBODY6} the \citet{2004MNRAS.353...87E} recipe is now used.
To test this, we carried out a simulation of model $128$kN with the \citet{2000MNRAS.315..543H}
prescription for stellar remnants, but we obtained a dissolution time
of $T_{diss}=23.0$ Gyr, which reduces the discrepancy by only
  about 30\% (see data for model 128kN in Table~\ref{parameter}).

  To assess stochastic effects (such as run-to-run variations) we also
  performed additional simulations of models $128$kN and $64$kN by
  evolving different numerical realizations of the same initial
  conditions, and by evolving the same realization in several
  independent simulations (as in
  \citetalias{2003MNRAS.340..227B}). Finally, we performed a
  simulation of model $128$kN in which the escapers were progressively
  removed (as in \citetalias{2003MNRAS.340..227B}), but again without
  any significant difference ($T_{diss}=22.9$ Gyr). 

None of these effects was able, individually, to account for the observed discrepancy. Therefore, we believe that the small but systematic discrepancy between our models without NS kicks and the corresponding ones in \citetalias{2003MNRAS.340..227B} results from a combination of all the effects mentioned above, and others which we have not studied, including possible differences in the way in which models are virialised and scaled in different codes.  As we shall show later (Sect.~\ref{sec:2modes}) the sensitivity of these runs to small effects is such that apparently trivial differences could have significant effects.

\section{Discussion}\label{Discussion}

\subsection{Two modes of star cluster dissolution}\label{sec:2modes}

We have found that the presence or absence of neutron star kicks, in the models we have studied, can change the lifetime of a star cluster
by a large factor.  We shall now try to interpret our results in the context of previous studies of star cluster dissolution mechanisms,
with the aim of understanding why it is that a process which affects such a small fraction of the mass can have such a dramatic effect.

We consider initially tidally filling, multi-mass models with stellar evolution.  Over the years, several numerical investigations have
shown that the dissolution time is strongly affected by two factors: the initial relaxation time and the initial concentration (represented
by the King parameter $W_0$).  In particular \citet{1990ApJ...351..121C} showed that, for a Salpeter-like IMF, their models with $W_0=1$ or $3$
dissolved quickly, in less than a Gyr, and without core collapse, while models with $W_0 = 7$ all entered core collapse, after about 10 Gyr or
longer.  Clusters with $W_0=3$ and a steeper IMF (and hence a longer time scale for mass loss by stellar evolution) could enter core collapse
before dissolution, provided that relaxation was fast enough.  Thus there is a tension between the time scales of stellar evolution and
relaxation, which plays out differently depending on the concentration. 

Recently \citet{2013ApJ...778..118W} noted that the clusters which
dissolve by the effects of stellar evolution lose their mass in a
qualitatively different way from those dominated by relaxation.

The former, as they approach dissolution, lose the last fraction of their mass (which may be substantial) extremely rapidly, whereas the latter lose mass at a rate which is steady, and sometimes even declining.  
Whitehead et al. also noted that the dividing line between the two modes of dissolution is quite sharp.  For that reason it would not be surprising if a very small effect, such as the loss or retention of NS, were to place a cluster in one mode of dissolution or the other.  

The two kinds of behaviour described by \citet{2013ApJ...778..118W}
are plainly visible in several previous studies of star cluster
evolution, such as \citet{2000ApJ...535..759T}, and they are visible
in Fig.~\ref{fig:MboverM0_128k} of the present Letter, where all
models with kicks end their evolution by losing mass precipitately
(except for the case N = $8$k), whereas the others lose mass at a more
moderate rate.  We refer to these two cases as ``jumping'' and
``skiing'', respectively\footnote{This unconventional terminology was
  coined by Simon Portegies Zwart in conversation with one of us (DCH)
  several years ago.  It vividly conveys the difference between skiing down a gentle slope and jumping off a cliff. We note that \citet{2013ApJ...778..118W} have conflated the two terms, with a different semantics.}. Fig.~\ref{fig:MboverM0_128k} also illustrates the point made by \citet{1990ApJ...351..121C}, i.e. the two modes of dissolution are characterised by the presence or absence of core collapse before dissolution. Indeed we see that the clusters with and without natal kicks (except for the case N = $8$k) lie on either side of the divide between the two modes.  

\begin{figure}
\includegraphics[trim=0 20 0 50, clip, width=0.48\textwidth]{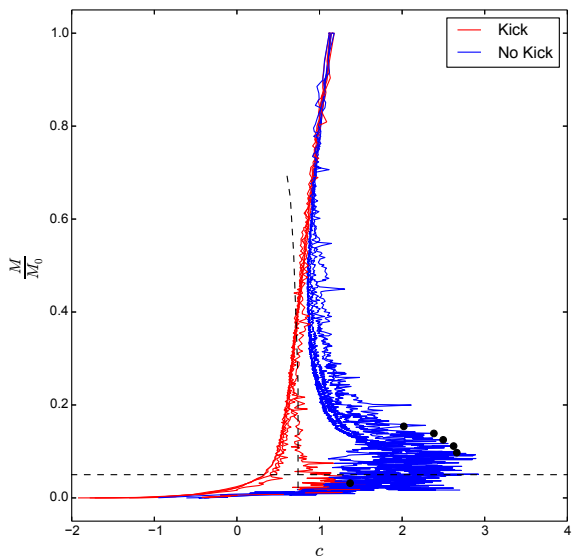}
\caption{The plot shows the total mass remaining in the cluster as a
  function of concentration,
 for the models depicted in Fig.~\ref{fig:MboverM0_128k}
with (red lines) and without (blue lines) NS kicks. The black
dots show the moment of core collapse, the black horizontal dashed
line at $M/M_0=0.05$ marks the formal dissolution condition, and the
black vertical dashed line denotes  
``Weinberg's cliff'' \citep[see
Fig.~\ref{fig:mmean} in][]{1993ASPC...48..689W}.
}\label{fig:weinberg10}
\end{figure}

\begin{figure}
\includegraphics[trim=0 20 0 50, clip, width=0.48\textwidth]{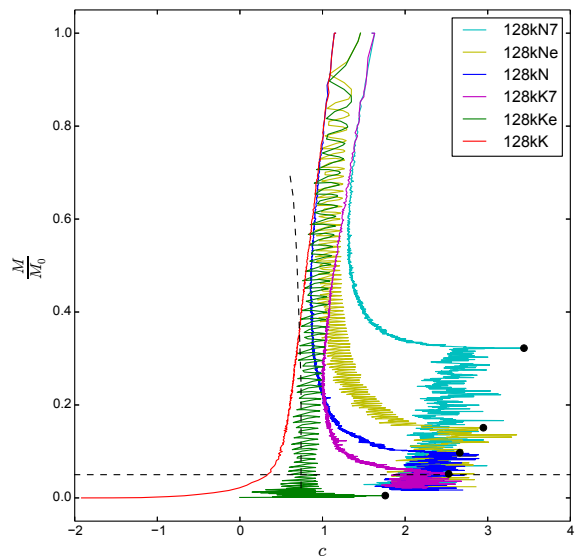}
\caption{As in Fig.~\ref{fig:weinberg10}, but presenting all models with N=$128$k. Note that model $128$kKe (green line) spans the region occupied by the separatrix, distinguishing ``skiing'' ($128$kN7, $128$kNe, $128$kN, $128$kK7) and ``jumping'' models ($128$kK).  }
\label{fig:weinberg_main}
\end{figure}

In order to visualise the transition between skiing and jumping models, it has been particularly instructive for us to take on the
point of view first suggested by \citet{1993ASPC...48..689W}, and to explore the evolution of our models in the plane defined by the
concentration (parameterized by $c=\log(r_J/r_c)$, where $r_c$ is the core radius) and the mass which remains bound to the system. In this
representation, a system which experiences exclusively stellar evolution effects would gradually lose mass, while reducing its
concentration due to the progressive expansion, giving rise to a track moving down in the plane and to the left  \citep[see Fig.~\ref{fig:mmean} in][]{1993ASPC...48..689W}. The tracks qualitatively resemble those of some of our models, as shown in Fig.~\ref{fig:weinberg10}.  These are four of the models with kicks, shown in red; and one of these (128kK) is also shown in Fig.~\ref{fig:weinberg_main}.

In Weinberg's treatment, dealing with the slow evolution of spherical equilibrium models, the tracks end when equilibrium is no longer possible; the tracks end at points along a curve, which is shown as a dashed near-vertical curve in these figures.  \nbody models can cross this
curve, but then lose mass on a dynamical time scale, explaining the jumping profile of the corresponding curves in Figs.~\ref{fig:MboverM0_128k} and \ref{fig:MboverM0_ECC_W7}.  Though its precise position may differ slightly when the simplifying assumptions of Weinberg's models are relaxed, we refer to this curve as ``Weinberg's cliff''.

In Weinberg's models, mass-loss is driven by stellar evolution only, and his results
should be applicable when this process dominates. When the effects of two-body relaxation are dominant, one of the natural consequences is the progressive increase of the central concentration, leading to core collapse. This results in a track oriented to the right-hand-side of the plane, behaviour which can be immediately recognised in the remaining models in Fig.~\ref{fig:weinberg10} and \ref{fig:weinberg_main}. It should not come as a surprise now that all long-lived, ``skiing'' models show signatures of core collapse, in contrast with short-lived, ``jumping'' models. 

These figures strongly suggest the existence of trajectories in which the two processes, stellar evolution and relaxation, are in a delicate balance overall, even though stellar evolution dominates early on and relaxation dominates thereafter. We have been particularly fortunate to have included in our survey two models whose evolution almost perfectly delimits a ``separatrix'' between ``skiing'' and ``jumping'' systems (see the innermost pair of red lines in Fig.~\ref{fig:weinberg10}, which correspond to models $8$kK and $16$kK). Even more strikingly, the model $128$kKe, despite the oscillations generated by the time-dependent tide, offers an excellent representation of the ``separatrix'' (see the green line in Fig.~\ref{fig:weinberg_main}).

Evidently, the models that we have studied lie close to the separatrix
dividing jumping models (which are dominated by stellar evolution,
lose mass rapidly at the end of their lives, and do not reach core
collapse) and skiing models (which are dominated by two-body relaxation, lose mass gently towards the end of their lives, and reach core collapse). If neutron stars are given no natal kick, as in model $128$kN, or the models of \citet{2003MNRAS.340..227B}, the trend to mass segregation and core collapse is accentuated, and the model moves across the separatrix into the domain of relaxation-dominated evolution. But we warn the reader against interpreting this as a general rule. Kicks were applied to both model 8kK and model 16kK (the innermost pair of red lines in Fig.~\ref{fig:weinberg10}), and they lie on opposite sides of the separatrix. The 8kK model, because of the low particle number and consequently smaller relaxation time, is 
sufficiently dominated by relaxation to lie in the skiing regime.

These considerations
do not immediately explain, however, why the lifetime should be so different as a factor of nearly four.  
But the example of models 32kK and 8kN, which lose mass in almost the same way until core collapse in the latter model (Fig.\ref{fig:MboverM0_128k}), shows that the effects of skiing and jumping lead to different lifetimes. Though the difference is only a factor 1.13 in this case, it seems plausible that the effect could be much bigger if the event which determines the mode of dissolution occurred very early in the lifetime of a model, e.g. the ejection of neutron stars. Furthermore, because our models lie so close to the separatrix between the two modes, it would not be surprising if very minor systematic differences in the initial conditions were to lead to significant systematic differences in the lifetime, as discussed in Sect.~3.2.

While this Letter has focused on kicks by neutron stars, the lesson to
learn is that apparently minor changes can have very large effects,
especially for clusters close to the transition between different
modes of dissolution.  Other factors which should be taken into
account include the presence and  properties of primordial binaries, variations in
the high-mass end of the IMF, and the degree of primordial mass
segregation, which influences both the importance of mass loss by
stellar evolution and the role of remnants, not only NS but also
stellar-mass black holes.
The importance of these factors depends on the location of the
dividing line between the two modes of dissolution that we have
discussed, which can be assessed only by means of appropriate numerical experiments.

\subsection{Conclusions}
We have presented evidence, based on $N$-body simulations of the evolution of initially tidally filling King models with stellar
evolution, that the presence or absence of NS natal velocity kicks can play a crucial role in the long-term survival of
model star clusters.  In particular we show that some of the basic models in the landmark study of \citet{2003MNRAS.340..227B} are especially sensitive to this effect, which can change their lifetime by almost a factor of four. We explain this finding by showing that the models lie close to a dividing line between (i) models which are dominated by the effects of mass-loss from stellar evolution, and whose evolution ends with a steepening rate of mass loss, and (ii) models whose dynamical evolution is dominated by two-body relaxation, which reach core collapse before dissolving, and do so with a gently decreasing rate of mass loss.

\section*{Acknowledgments}\label{acknowledgments}
We thank Mark Gieles and Simon Portegies Zwart for useful comments and valuable discussions and an anonymous referee for constructive comments. 
The simulations were carried out on GeForce GTX 780 graphics cards at University of Surrey and we thank Dave Munro for the hardware support. FC acknowledges support from the European Research Council (ERC-StG-335936, CLUSTERS), and ALV from the Royal Commission for the Exhibition of 1851. This work was initiated during the 2014 International Summer Institute for Modeling in Astrophysics, hosted by CITA at the University of Toronto. We are grateful to Pascale Garaud for its organisation, for financial support and, together with the other participants, for the stimulating research environment.

\bibliographystyle{mn2e}

\label{lastpage}

\end{document}